\documentclass[%
superscriptaddress,
preprint,
 amsmath,amssymb,
 aps,
]{revtex4-2}

\usepackage{graphicx}
\usepackage{dcolumn}
\usepackage{bm}
\usepackage{setspace}

\newcommand{\UB}{\mathbf U_\text{B}}
\newcommand{\UN}{\mathbf U_\text{N}}
\newcommand{\Ap}{{A$'~$}}
\newcommand{\App}{{A$''~$}}
\newcommand{\Bp}{{B$'~$}}
\newcommand{\Cp}{{C$'~$}}

\begin{document}


\title{Origins of the transformability of Nickel-Titanium shape memory alloys}%

\author{Xian Chen} \thanks{Corresponding author: Xian Chen xianchen@ust.hk}
\affiliation{Mechanical and Aerospace Engineering, Hong Kong University of Science and Technology, Hong Kong}
\author{Colin Ophus}%
\affiliation{National Center for Electron Microscopy
Molecular Foundry, Lawrence Berkeley National Laboratory, California, United States}
\author{Chengyu Song}
\affiliation{National Center for Electron Microscopy
Molecular Foundry, Lawrence Berkeley National Laboratory, California, United States}
\author{Jim Ciston}
\affiliation{National Center for Electron Microscopy
Molecular Foundry, Lawrence Berkeley National Laboratory, California, United States}
\author{Sambit Das}
\affiliation{
 Mechanical Engineering, University of Michigan, Ann Arbor, Michigan, United States
}%
\author{Yintao Song}
\affiliation{Aerospace Engineering and Mechanics, University of Minnesota, Minnesota, United States}
\author{Yuriy Chumlyakov}
\affiliation{Tomsk State University Siberian Physical-Technical Institute
Tomsk, Russia}
\author{Andrew M. Minor}
\affiliation{National Center for Electron Microscopy
Molecular Foundry, Lawrence Berkeley National Laboratory, California, United States}
\affiliation{University of California, Berkeley, California, United States}
\author{Vikram Gavini}
\affiliation{
 Mechanical Engineering, University of Michigan, Ann Arbor, Michigan, United States
}
\author{Richard D. James}\thanks{Corresponding author: Richard D. James james@umn.edu}
\affiliation{Aerospace Engineering and Mechanics, University of Minnesota, Minnesota, United States}

\date{\today}

\begin{abstract}
The near equiatomic NiTi alloy is the most successful shape memory alloy by a large margin. It is widely and increasingly used in biomedical devices. Yet, despite having a repeatable superelastic effect and excellent shape-memory, NiTi is very far from satisfying the conditions that characterize the most reversible phase transforming materials. Thus, the scientific reasons underlying its vast success present an enigma. In this work, we perform rigorous mathematical derivation and accurate DFT calculation of transformation mechanisms to seek previously unrecognized twin-like defects that we term involution domains, and we observe them in real space in NiTi by the aberration-corrected scanning transmission electron microscopy. Involution domains lead to an additional 216 compatible interfaces between phases in NiTi, and we theorize that this feature contributes importantly to its reliability. They are expected to arise in other transformations and to alter the conventional interpretation of the mechanism of the martensitic transformation. 
\end{abstract}

\maketitle

\section{Introduction}
The near equiatomic NiTi alloy is a key element in a vast array of medical devices, including stents, guidewires, embolic filters, dental arch wires, bone implants and microforceps. \cite{petrini2011} NiTi alloys also are essential to current and emerging designs of brain stents \cite{wakhloo} and devices for deep brain stimulation \cite{bechtold}. These medical applications depend critically on the ability of the NiTi alloy to pass reversibly back and forth through its big symmetry-breaking phase transformation. 

Many authors have presented explanations for the success of NiTi. The desirable role of precipitates in slightly Ni rich NiTi is widely accepted \cite{Miyazaki_1989}. These strengthen the high-temperature austenite phase, and mitigate against dislocation motion during transformation. Some prior plastic deformation has a similar effect \cite{Miyazaki_1989}. Grain size also plays a role \cite{gu2018phase,yin2016} with fine, but not too fine (i.e. around 50nm), grain size as ideal. The need to avoid contamination by carbide and nitride particles that become sites for fatigue-crack initiation is widely appreciated \cite{tabanli}. In this report we present an alternative hypothesis for the reliability of NiTi based on the observation that it allows for a plethora of previously unrecognized interfaces between phases. 

Recently, the elimination of stressed transition layers \cite{Waitz_2005, Delville_2010} between phases has been shown to profoundly influence the reversibility and hysteresis of phase-transforming materials \cite{Zarnetta2010, Song_2013, Chluba_2015}. The most highly reversible alloys measured by the absence of degradation of properties under cycling now satisfy such conditions of \emph{supercompatibility} \cite{Chen_2013}. However, the near equiatomic NiTi alloys are far from satisfying any known conditions of supercompatibility. We demonstrate in work, however, that NiTi satisfies a certain nongeneric involution relation, leading to many additional nonstandard compatible interfaces. An involution is a mapping f of a domain to itself that, when applied twice, gives the identity: $f(f(x)) = x$.

\section{Result and discussion}
\subsection{Involution domains}

\begin{figure*}[ht]
\centering
\includegraphics[width=0.95\textwidth]{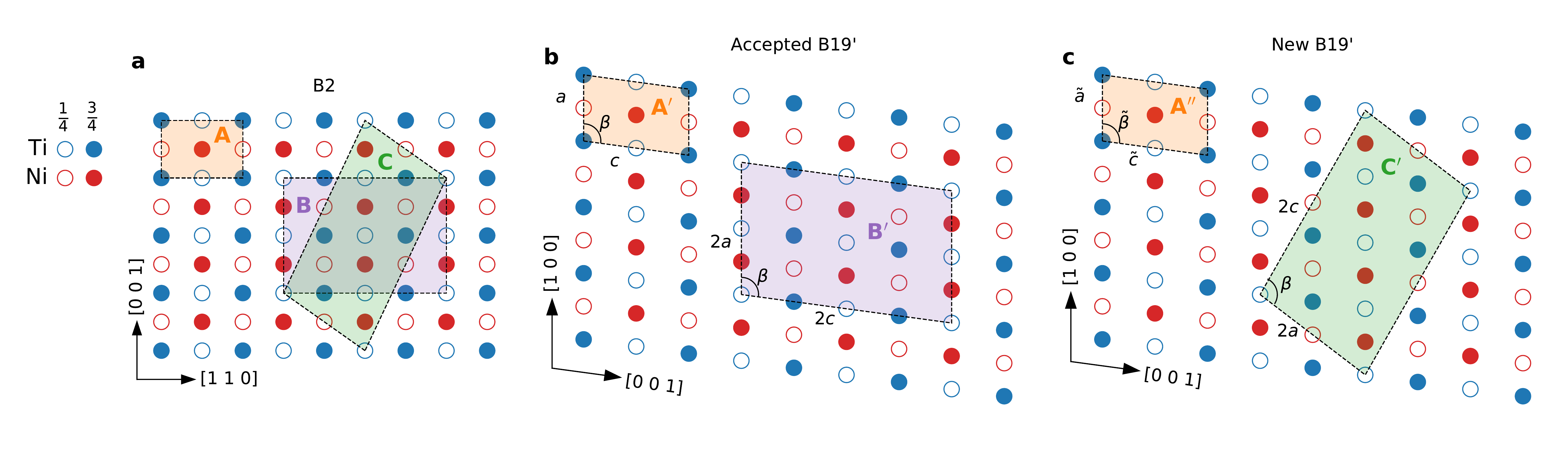}
\caption{(a) The $(1 \bar1 0)_\text{B2}$ projection of B2 lattice corresponding to the $(010)_\text{B19'}$ projection of 
deformed lattice by (b) accepted mechanism and (c) new mechanism respectively. Here we only plot the atom positions 
under the homogeneous deformation given by $\UB$ and $\UN$.}\label{fig1:atoms}
\end{figure*}

To explain the origins of this involution, we revisit accepted ideas of the crystal structure of NiTi. Following a period of debate in the 1960s \cite{Miyazaki_1989, Purdy_1965,Marcinkowski_1968,Otsuka_1971, Otsuka_1977, Knowles_1980, Kudoh_1985, Miyazaki_1984, Otsuka_2005}, the B19'  crystal structure of the martensite phase of NiTi and the associated transformation mechanism - which atom goes where - are now well accepted. B19'  is monoclinic (P21/m symmetry) with a 4-atom unit cell \cite{Kudoh_1985}. This phase is deformed from the B2 phase (Pm3m symmetry) by a basal shear on the $(110)_\text{B2}$ plane along a $[001]_\text{B2}$ direction \cite{Otsuka_1971} which yields a Bain-like strain \cite{Knowles_1980, Miyazaki_1984, Bain_1924}, illustrated as Fig. \ref{fig1:atoms}a, b. During this transformation, the 4-atom unit cell of the B2 sublattice with a basis of $[001]$, $[1\bar10]$ and $[110]$ (marked as the cell A in Fig. \ref{fig1:atoms}a) becomes the 4-atom unit cell of the product phase B19' (marked as the cell A’ in Fig. \ref{fig1:atoms}b). This 4-atom unit cell is widely accepted as the primitive unit cell of the B19'  phase, but shuffling of atoms within this unit cell (consistent with this periodicity) occurs. The sequence of all $(hh0)_\text{B2}$ layers does not shear, so the axis $[110]_\text{B2}$ simultaneously becomes the monoclinic 2-fold axis $(010)_\text{B19'}$ for the accepted B19' phase. This is known as Bain correspondence \cite{Knowles_1980, Bain_1924}. 

Recently, a variety of theoretical and algorithmic methods have emerged for the interpretation of experimental data in materials science, especially for phase transformations \cite{Ichiro_15, Lee_2011}. One such method is an algorithm (StrucTrans \cite{Chen_2016a}) for determining the lattice correspondence with the smallest strain, among all correspondences that map a sublattice of the parent phase to the lattice of the product phase. The strain here is determined from the lattice correspondence by the Cauchy-Born rule \cite{Ericksen_2008, bhatta2004}. This is a rigorous algorithm that is mathematically guaranteed to converge in a finite number of steps to the linear transformation between these lattices with the smallest strain. This algorithm revealed an unexpected new transformation mechanism explained below, which led us to seek involution domains. 

Is the Bain transformation the most likely, based on a smallest strain criterion for deforming the unit cell A to the unit cell \Ap (Fig. \ref{fig1:atoms}a,b)? The StrucTrans algorithm addresses exactly this question by searching for the lattice correspondence that minimizes the transformation strain defined by $\frac{1}{3}\sqrt{\sum_{i=1}^3 \epsilon_i^2}$ where $\epsilon_1$, $\epsilon_2$ and $\epsilon_3$  are the principal strains. The result indeed confirms that the Bain correspondence gives the smallest strain, and there are no near competitors. 

However, by testing other sublattices we noticed a peculiar feature of the transformation in NiTi which is reflected in the similarity of the purple and green cells of Fig. \ref{fig1:atoms}b, c. Using the most widely accepted values of lattice parameters for NiTi of Kudoh et al. \cite{Kudoh_1985}, there exists a smaller transformation strain than the Bain strain for the green sublattice of B19'. In this case, it is possible that the crystal may undergo a new transformation that deforms the sublattice C (green cell in Fig. \ref{fig1:atoms}c), then shuffles the atoms inside the sublattice unit cell to finally achieve a stable new B19' structure. In practice, these two processes - homogeneous deformation of the cell and atomic shuffling within the cell - would typically occur simultaneously. 

A concrete example is based on the reported lattice parameters, $a_0= 3.015\AA$ \cite{Otsuka_1971} for B2 and $a=2.898\AA$, $b=4.108\AA$, $c=4.646\AA$ and $\beta = 97.78^\circ$\cite{Kudoh_1985} for B19’. The new mechanism deforms the green cell C of Fig. \ref{fig1:atoms}a to the monoclinic sublattice \Cp in Fig. \ref{fig1:atoms}c with transformation strain $0.0472 \pm 0.0001$. On the other hand, the Bain mechanism has the strain $0.0475 \pm 0.0005$. The deformed sublattice unit cell \Bp in Fig. \ref{fig1:atoms}b is a rotation and translation of the sublattice unit cell \Cp in Fig. \ref{fig1:atoms}c.

The key difference between the new and the Bain mechanism is their transformation stretch tensors. Considering the unit cells A, \Ap and \App in Fig. \ref{fig1:atoms}, there are two transformation stretch tensors, the square of which can be expressed analytically in terms of lattice parameters as
\[
\UB^2 = \begin{bmatrix}\alpha & \delta & \eta\\ \delta & \alpha & \eta\\ \eta & \eta & \gamma\end{bmatrix} \text{ and }
\UN^2 = \begin{bmatrix}\tilde{\alpha} & \tilde{\delta} & \tilde{\eta} \\ \tilde{\delta} & \tilde{\alpha} & \tilde{\eta} \\ \tilde{\eta} & \tilde{\eta} & \tilde{\gamma}\end{bmatrix}, 
\]
where
$\alpha = \frac{b^2 + c^2}{4 a_0^2}$, $\delta = \frac{-b^2 + c^2}{4 a_0^2}$, $\eta = \frac{a c \cos \beta}{2 a_0^2}$ and $\tilde{\alpha} = \frac{9 a^2 + 4 b^2 + c^2 + 6 a c \cos \beta}{16 a_0^2}$, $\tilde{\delta} = \frac{9 a^2 - 4 b^2 + c^2 + 6 a c \cos \beta}{16 a_0^2}$, $\tilde{\eta} = \frac{3 a^2 - c^2 - 2 a c \cos \beta}{8 a_0^2}$, $\tilde{\gamma} = \frac{a^2 + c^2 - 2 a c \cos \beta}{4 a_0^2}$. 

They deform A to \Ap and \App, respectively. Since \Ap$\neq$ \App, then $\UB \neq \UN$, and $\UN$ is also not symmetry-related to $\UB$. However, $\UN$ is not just an arbitrary perturbation of $\UB$. In fact they are related through a strong crystallographic restriction: after a homogeneous deformation, the sublattice \Bp of B19' has the same periodicity as the sublattice \Cp(Figs. \ref{fig1:atoms}b, c) although they orient differently. Let 
\[
g_1=a_0 [002]_\text{B2},\ g_2=a_0 [1\bar10]_\text{B2},\ g_3=a_0 [220]_\text{B2}
\] be the basis of the sublattice B, and 
\[
\tilde{g}_1 = a_0 [111]_\text{B2},\ \tilde{g}_2 = a_0 [1\bar10]_\text{B2},\ \tilde{g}_3 = a_0[11\bar3]_\text{B2}
\]
be the basis of the sublattice C. After phase transformation, the bases of the deformed cells \Bp and \Cp are calculated as $m_i = \mathbf R_\text{B} \UB g_i$ and $\tilde{m}_i = \mathbf R_\text{N} \UN \tilde{g}_i$ for some rigid rotations $\mathbf R_\text{B}$ and $\mathbf R_\text{N}$. The bases of the deformed cells \Bp and \Cp satisfy the relationship
\begin{equation}\label{eq:metric}
    m_i \cdot m_j = \tilde{m}_i \cdot \tilde{m}_j.
\end{equation}
Let $(\tilde{a}, b, \tilde{c}, \tilde{\beta})$ denote the lattice parameters of the cell \App deformed by the new mechanism as noted in Fig. \ref{fig1:atoms}c. Obviously, they are slightly different from those of the cell \Ap deformed by the Bain mechanism, given by Fig. \ref{fig1:atoms}b. By the geometrical restriction \eqref{eq:metric}, the lattice parameters describing the cells \Ap and \App must satisfy 
\begin{align*}
&4 \tilde a^2 = a^2 + c^2 - 2 a c \cos \beta \tag{C1} \label{c1}\\
&4 \tilde c^2 = 9 a^2 + c^2 + 6 a c \cos \beta \tag{C2} \label{c2}\\
&4 \tilde a \tilde c \cos \tilde \beta = -3 a^2 + c^2 + 2 a c \cos \beta \tag{C3}\label{c3}.
\end{align*}
Note that, the above equations are independent of $a_0$ (\emph{i.e.} the cubic lattice parameter of austenite phase). For any monoclinic lattice with the primitive cell having the lattice parameters $(a,b,c,\beta)$, equations \eqref{c1} to \eqref{c3} calculate a set of new monoclinic lattice parameters $(\tilde{a}, b, \tilde{c}, \tilde{\beta})$. In addition, \eqref{c1} to \eqref{c3} is an exact involution on $a, c, \beta$ space (One can add $b$ since it is unchanged). Under mild conditions, involutions have fixed points and that is the case here: there is a two-dimensional surface in $a$,$c$,$\beta$ - space where \eqref{c1} to \eqref{c3} satisfies $\tilde a = a$, $\tilde c = c$ and $\tilde \beta = \beta$. We believe that a key to understanding the reversibility of NiTi is the fact that its lattice parameters lie extremely close to this surface. For example using the reported $a = 2.898 \AA$, $c = 4.646 \AA$ and $\beta=97.78^\circ$ \cite{Kudoh_1985} for \Ap, the values of $\tilde a$, $\tilde c$ and $\tilde \beta$ are $2.8995$\AA, $4.6431$\AA, and $97.743^\circ$.

\subsection{Theoretical exploration for the involuted structure}

To the best of our knowledge, the most accurate way to determine lattice parameters of crystalline solids is X-ray diffraction. The accuracy depends on the quality of samples, X-ray source and geometrical factors of the facility. The best achievable accuracy for solving a monoclinic structure is about 0.0001Å \cite{Kudoh_1985}, which is not sufficient to distinguish the new B19' structure from the accepted one. We simulated the X-ray powder diffraction patterns (Fig. S6 in Supplementary Information) for the 4-atom A' unit cell and the 16-atom C' unit cell respectively. Despite of different indexing labels, the two patterns are almost identical to each other.

We have therefore conducted density functional theory (DFT) calculations to investigate the viability of the involuted B19' structure. Many DFT studies have been conducted over the past two decades on the low temperature phase of NiTi, but the lowest energy crystal structure found was not in agreement with experiments \cite{Huang_2003, Wagner_2008, Hatcher_2009, Lawson_2016}. Problematically, structures close to the accepted crystal structure of martensite were not stable with respect to small perturbations. Here, we employ recently developed optimized norm-conserving Vanderbilt pseudopotentials \cite{ONCV_2013}, and obtain a stable monoclinic phase with P2$_1$/m symmetry that is consistent with experiments. 

In particular, we consider both 4-atom unit cell (i.e. \Ap or \App) and the 16-atom unit cell (i.e. \Bp or \Cp). The initial lattice parameters used for geometry optimization for both the austenite and martensite phases were adopted from \cite{Kudoh_1985}. The starting Wyckoff atomic positions are rational, and calculated from the B2 lattice using the lattice correspondences given by Bain and new mechanisms, respectively. Prior to the optimization of the geometry of those structures, the atomic positions were randomly perturbed to avoid constraining the optimization to a manifold. The results obtained were independent of the perturbation. Upon geometric relaxation of both cell parameters and ionic positions, we obtain: (i) for the 4-atom unit cell, P2$_1$/m symmetry with $a = 2.944$\AA, $b = 4.028$\AA, $c = 4.805$\AA and $\beta = 103.078^\circ$; (ii) for the 16-atom unit cell, P2$_1$/m symmetry with $2a = 6.179$\AA, $b = 4.027$\AA, $2c = 9.05$\AA and $\beta = 99.593^\circ$. The energies of both (i) and (ii) relative to the B2 phase are the same up to numerical precision (Table S2-4 in Supplementary Information). Interestingly, the lattice parameters of the 4-atom unit cell are related to those of the 16-atom unit cell via the involution \eqref{c1} to \eqref{c3}, thus providing evidence for the new mechanism. We note that lattice parameters deviate from experiment by a few percent, which is consistent with the accuracy expected from DFT \cite{Lejaeghere_2016}. 

We remark that the feature of NiTi of satisfying an involution relation is unusual.  In contrast to NiTi, a typical $\beta$-phase (e.g., non-modulated) alloy CuAlZn \cite{chakravorty1977} has monoclinic structural parameters $a = 4.553$\AA, $b = 5.452$\AA, $c = 4.33$\AA and $\beta = 87.5^\circ$. After the involution transformation  \eqref{c1} to \eqref{c3} is applied to this phase, we get $\tilde a = 3.0724$\AA, $\tilde c = 7.2539$\AA and $\beta = 117.904^\circ$. Clearly, the involuted lattice has an unrealistically large lattice distortion and is highly incompatible with the austenite phase. In contrast, the compatible involution domains are possible in NiTi.

Due to the nature of involution, a martensite variant and an involuted variant can alternate just like a twinned structure (from equations S6-S10 in SI). This new kind of interface differs from a conventional twin since the atom positions in the unit cells across the interface lacks mirror symmetry. We called this new family of twin-like microstructures involution domains. By solving the compatibility equation \cite{Ball_1987}, we also find that laminates of involution domains are compatible with austenite across a habit plane with a volume fraction 50\%, which distinguishes them from the accepted twins having twin ratio either 68\% for type I twin or 73\% for type II twin \cite{Knowles_1980}. We used this fact to locate them in the electron microscope.

\subsection{Experimental exploration for the involution domains}

The search for involution domains between lattices deformed by the accepted and new mechanisms in NiTi was carried out using aberration-corrected scanning transmission electron microscopy (STEM). We use the high angle annular dark field (HAADF) images to directly study the morphology of the interface and atomic positions in real space \cite{nellist2011}. Since the lattice parameters and corresponding theoretical involuted ones are too close for NiTi, i.e. they are identical up to the 3rd decimal place, diffraction based probes are not sufficient to distinguish the underlying periodicities between the two reciprocal lattices \cite{nellist2011}. By direct observation of atomic positions using HAAD-STEM, we can image the interface and the neighboring atom shuffles to compare with the involution domain in real space. 

\begin{figure}[ht]
    \centering
    \includegraphics[width=0.45\textwidth]{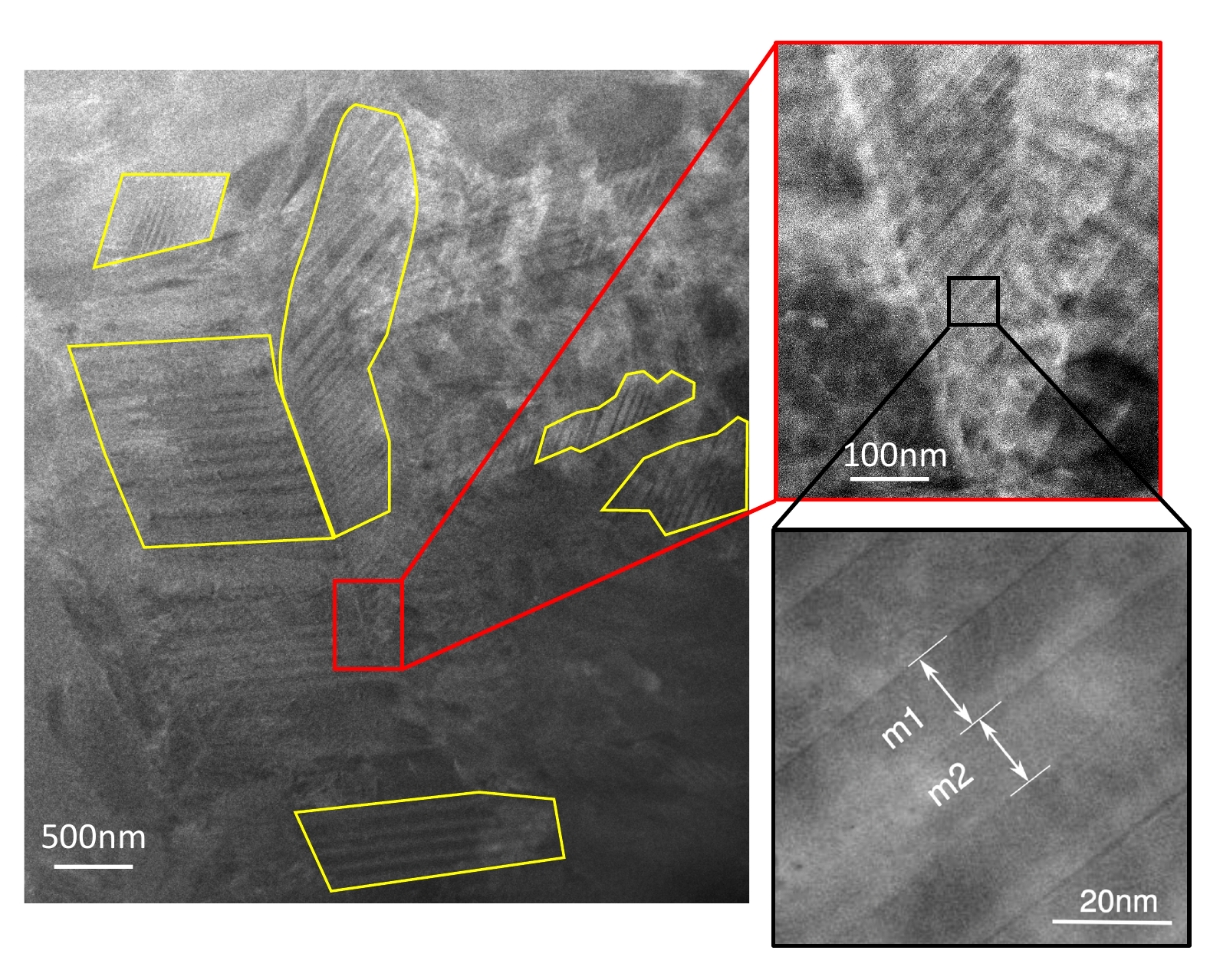}
    \caption{Low magnification STEM images showing the martensitic morphology in NiTi. The regions outlined by yellow lines correspond to classical type I/II twins of NiTi with volume fraction around 70\%. The domain morphology in the red box shows twin-like laminates with the ratio of width about 1:1. The neighboring bands are labeled as m1 and m2. }
    \label{fig:low_mag}
\end{figure}

We used a near equiatomic NiTi specimen, which was synthesized and heat treated to reduce the internal stress at room temperature. The electron transparent foil cut from the bulk martensite was thinned by electrolytic polishing to minimize the surface deformation. The austenite and martensite start/finish temperatures of the specimen were $88^\circ$C/$118^\circ$C and $78^\circ$C/$40^\circ$C (Supplementary Information). The R-phase transformation was not observed.

To search for the predicted twin-like interface, we tilted the foil so that the e-beam was as perpendicular to the involution domain normal as possible. In that case, we would expect the laminate morphology at relatively low magnification. In Fig. \ref{fig:low_mag}, we observed many martensitic twins and twin-like microstructures without obvious appearance of dislocations and large distortions. We outlined the typical twin morphologies in yellow whose volume fractions are nearly 70\% that agrees with the reported type I/II twins in literature \cite{Knowles_1980}. We also found in Fig. \ref{fig:low_mag} a small region of about $100 \times 100$nm$^2$, marked in a red box, showing a series of laminated structures with about equal width. It is distinct from the usual Type I and Type II twin structures at low magnification. As calculated by the compatibility equations \cite{Song_2013,Chen_2013} for non-generic rank-one domains, the twin-like bands have the volume fraction 50\% and the interface between them is straight and sharp, which agrees with the observed morphology very well. 

\begin{figure*}[ht]
    \centering
    \includegraphics[width=0.8\textwidth]{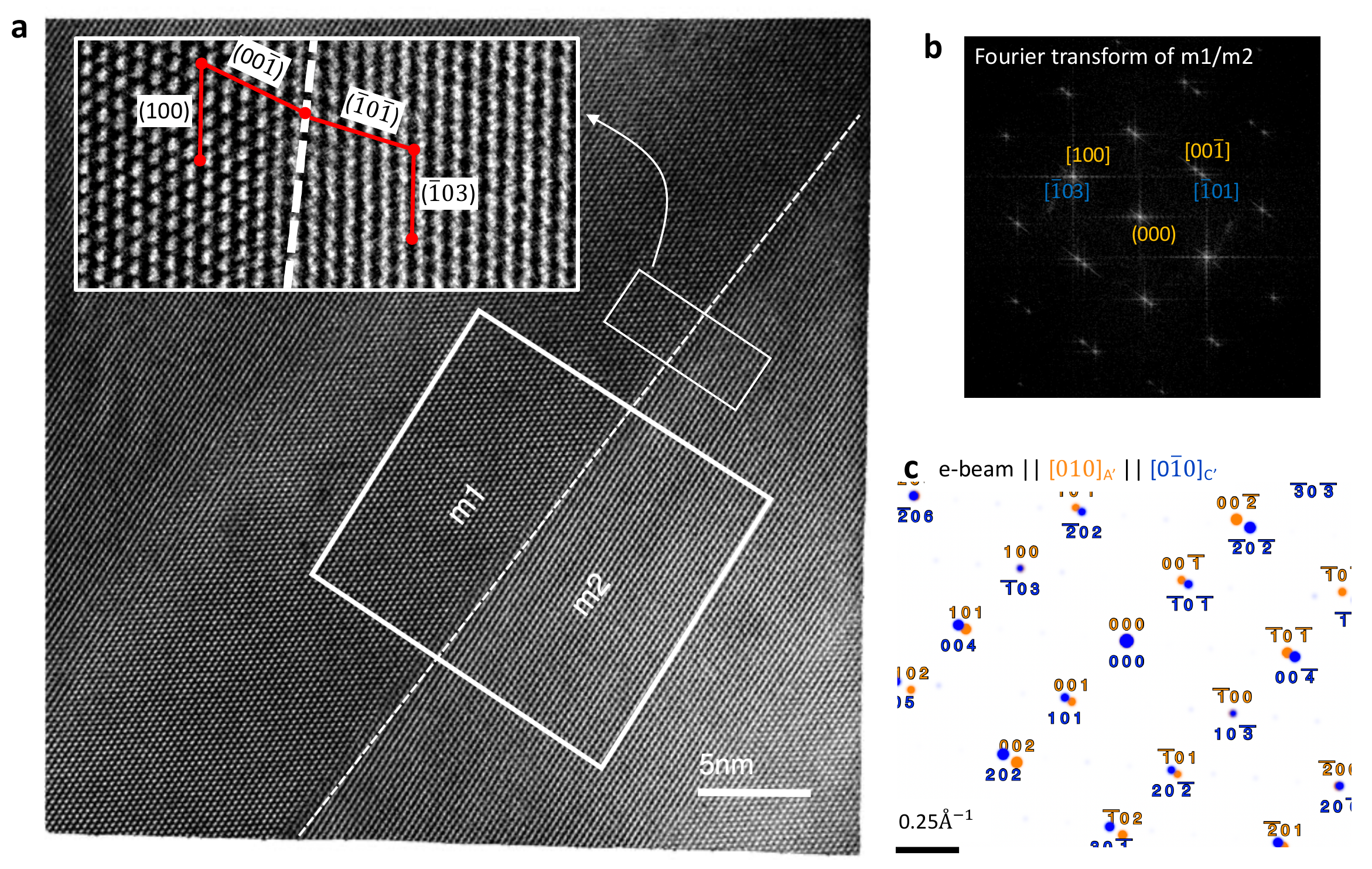}
    \caption{(a) The high resolution STEM image in the same area showing the atomic structures near the interface. (b) The Fourier transform of the STEM image within the white box. (c) The theoretical diffraction patterns of the 4-atom Bain cell (orange dots) and the 16-atom new cell (blue dots) along $[010]_\text{B19'}$ direction.}
    \label{fig:high_res}
\end{figure*}

The atomic columns are captured by double-axes tilting under the drift-corrected high resolution HAADF mode shown in Fig. \ref{fig:high_res}a. The Fourier transform of the white boxed area in Fig. \ref{fig:high_res}a is presented in Fig. \ref{fig:high_res}b. The Fourier transform implies a twin-like feature in reciprocal space, but the atomic arrangement across the interface are not mirror-related. We used the lattice parameters of the 4-atom cell (\Ap in Fig. \ref{fig1:atoms}b) and the 16-atom cell (\Cp in Fig. \ref{fig1:atoms}c) to simulate the diffraction patterns (Fig. \ref{fig:high_res}c) in $[010]_\text{B19'}$ zone axis. We found that the diffraction patterns capture the main periodicity revealed by Fourier transform. From the indexing, the crystallographic plane (100) of \Ap possibly parallels to $(\bar103)$ of \Cp. In real space, such a parallelism corresponds to the stack of atoms along the vertical lines plotted in Fig. \ref{fig1:atoms}b and c. Since $\UB$ and $\UN$ deform the B2 lattices under different lattice correspondences, the atomic shuffles in deformed B19' across the involution interface would be different from each other. Such subtle differences are captured in real space image for the interface, as the inset of Fig. \ref{fig:high_res}a.

\begin{figure}[ht]
\centering
\includegraphics[width=0.45\textwidth]{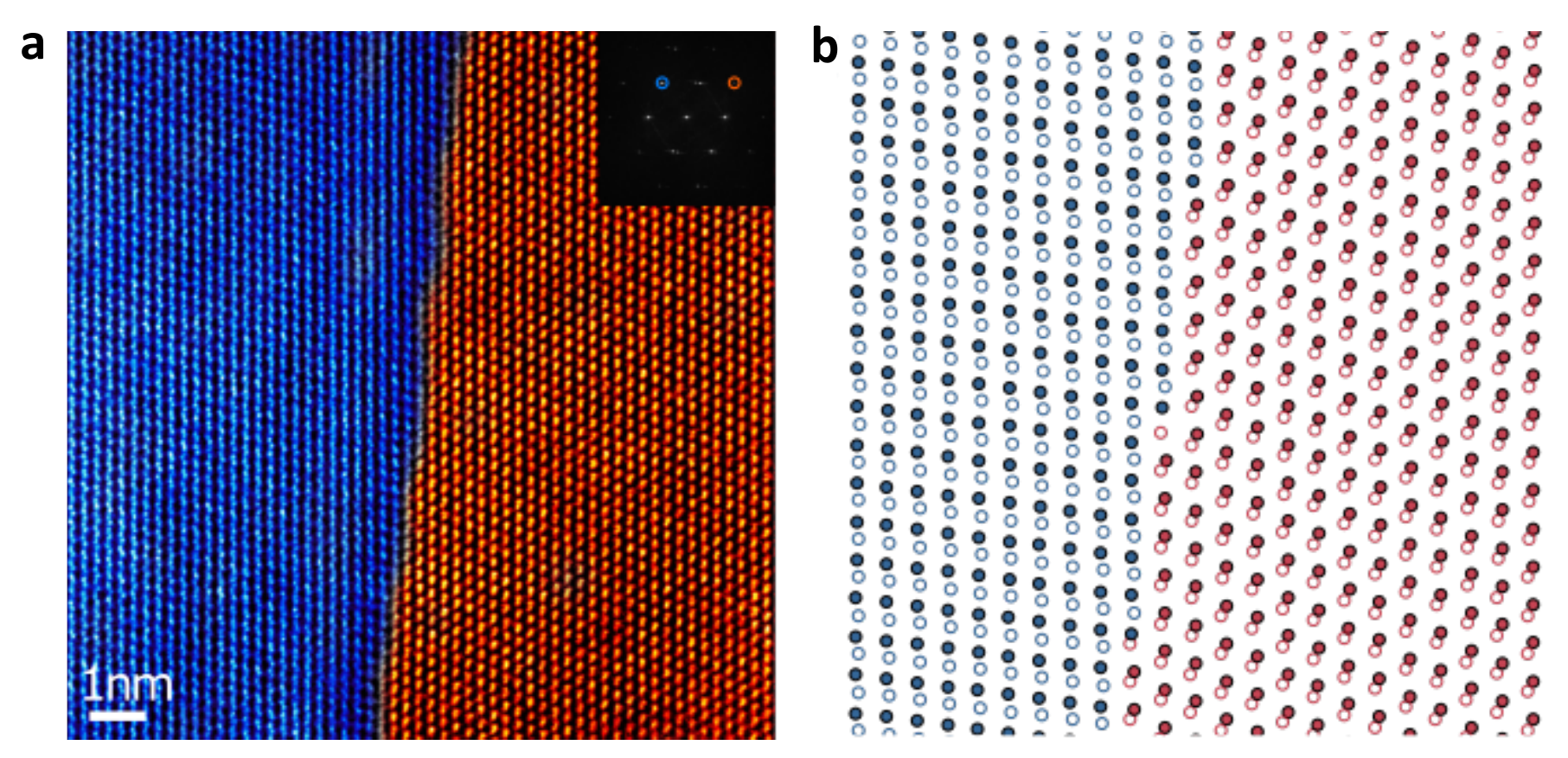}
\caption{Atomic structure of involution domains observed in NiTi. (a) STEM image of the atomic positions, where both lattices are color coded by Bragg filtering from the peaks shown in the image Fourier Transform inset in the upper right corner. (b) Calculated involution domain interface with proper atomic shuffles by the mechanics theory of compatibility \cite{Ball_1987}. The hollow and solid circles correspond to the atomic layers along $[010]_\text{B19’}$ whose heights differ by $\frac{1}{2}$ of the periodicity. }
\label{fig:interface}
\end{figure}

Fig. \ref{fig:interface}a shows the enlarged view of the boxed region M1 and M2 in Fig. \ref{fig:high_res}a. Using a Bragg peak filter (see Methods in Supplementary Information), the interface is resolved between the blue and orange lattices. The irrational morphology of the sharp interface shown in Fig. \ref{fig:interface}a rules out the possibility of Type I and Compound twins. It is unlikely to be the Type II twin interface either, because all possible Type II twins that can form compatible austenite/martensite interfaces should have the twinning volume ratio $0.27: 0.73$ \cite{Knowles_1980}. In contrast, the concept of an involution domain explains the irrationality, asymmetry and $1:1$ volume twinning volume ratio in Fig. \ref{fig:low_mag} and \ref{fig:high_res}. The near perfect piecewise linearity of the deformation as indicated by the Fourier transform in Fig. \ref{fig:high_res}b further supports the interpretation of the interface observed in Fig. \ref{fig:high_res} as an involution domain rather than, for example, an arbitrary stressed region arising from a nearby defect or inclusion. 

\begin{widetext}
\begin{center}
\begin{table}
\centering
\caption{Lattice parameters of the B19' structure (4-atom primitive unit cell) and the B2 austenite structure of the NiTi used for STEM.}
\label{tab1}
\begin{tabular}{ccccc}
\hline
 & $a (\AA)$ & $b (\AA)$ & $c (\AA)$ & $\beta (^\circ)$ \\
\hline
B2 & $3.0179\pm0.0005$ & $3.0179\pm0.0005$ & $3.0179\pm0.0005$ & 90\\
\hline
measured B19' & $2.8785\pm0.0004$ & $4.1106\pm0.0002$ & $4.6189\pm0.0001$ & $97.171\pm0.006$\\
\hline
calculated B19' & $2.8696\pm0.0003$&$4.1106\pm0.0002$ & $4.6354\pm0.0007$ & $97.39\pm0.01$\\
\hline
\end{tabular}
\end{table}
\end{center}
\end{widetext}

To verify the normal of this observed interface by theory, we measured the lattice parameters of B2 and B19' for this TEM foil (Table \ref{tab1}). The measured parameters in Table \ref{tab1} are slightly different from the reported ones \cite{Kudoh_1985}, which might be attributed to a thin foil effect of the TEM specimen. Using these measured parameters of B19’, \eqref{c1} to \eqref{c3} give the lattice parameters $\tilde a$, $\tilde c$ and $\tilde \beta$ by the new transformation mechanism, listed as the calculated B19' in Table \ref{tab1}. The two sets of lattice parameters are quite close, consistent with the transformation strains predicted by the StrucTrans algorithm. Solving the equations of compatibility \cite{Ball_1987} between the two strains, the calculated interface is $n = (1,1,-0.015)$ written in the orthogonal B2 basis (Supplementary Information). The lattices across the theoretical involution interface with proper atomic shuffles are plotted in Fig. \ref{fig:interface}b. The calculated irrational interface agrees well with the experimental observations, in support of our conjecture for the existence of the new transformation mechanism and the involution domains. 

\section{Conclusion and prospective}

A widely accepted idea in the study of martensitic phase transformations is that the number and types of compatible interfaces plays a key role in reversibility of the transformation, which is usually quantified by the repeatability of the superelastic strain. For example, a reversible shape memory effect has not been documented in any polycrystal that undergoes a tetragonal-to-orthorhombic transformation (only 2 variants) or any alloy that does not satisfy the conditions of the crystallographic theory of martensite \cite{lieberman_57} (some martensitic steels). In recent years this accepted idea has been extended to encompass special non-generic relations between lattice parameters at which unstressed interfaces between austenite and martensite become possible (``$\lambda_2 = 1$'' \cite{Zarnetta2010}), or many such interfaces become possible (“cofactor conditions” (\cite{Song_2013, Chluba_2015})). Currently, the alloys exhibiting the greatest resistance to the functional fatigue satisfy the latter. NiTi is not anywhere near to satisfying these conditions. Thus, aside from ancillary considerations such as its early-accepted biocompatibility, good strength and corrosion resistance \cite{Duerig_1999}, the success of its broad use in superelastic applications is a puzzle. Certainly, a large body of work has gone into optimizing its synthesis and processing, by the development of synthesis routes that yield exceptionally clean material without oxides, nitrides, etc., and also heat treatments that optimize the types, shapes, and sizes of precipitates that mitigate against fatigue mechanisms \cite{Miyazaki_1989}. In this work, we offer an additional reason for its success: it has in fact two monoclinic martensitic phases and 24 variants (instead of the accepted 12), which offers an additional 216 (\emph{i.e.} 192 twins + 24 non-generic domains by involution) compatible interfaces among new and accepted variants. The new defect -- involution domain -- underlies a new paradigm for the study of microstructure-property relation not only for the NiTi binary alloy, but also for the alloy systems with B19' structure (non-modulated monoclinic structure). For example, the nearby ternary NiTiNb systems with $9\sim 10$at.\% Nb contents potentially would have the involution domains because its lattice parameters are quite similar to binary NiTi \cite{shi2014}, corresponding to large thermomechanical hysteresis during cyclic loading processes. The discovery of the involution domains enables the defect engineering in general for monoclinic symmetry metals and alloys. 

\begin{acknowledgements}
XC acknowledges the financial support of the HK Research Grants Council through the General Research Fund under Grants 16207017 and 16201019. CO and JC acknowledge additional support from the US Department of Energy Early Career Research Program. RDJ was supported by NSF (DMREF-1629026), ONR (N00014-18-1- 2766), MURI (FA9550-18-1-0095, FA9550-16-1-0566) and a Vannevar Bush Faculty Fellowship. Work at the Molecular Foundry and Advanced Light Source was supported by the Office of Science, Office of Basic Energy Sciences, of the U.S. Department of Energy under Contract No. DE-AC02-05CH11231.
\end{acknowledgements}

\bibliography{niti_ref}

\end{document}